\documentclass[aps,amsmath,twocolumn,amssymb,floatfix,nofootinbib,longbibliography]{revtex4-1}
\pdfoutput=1
\usepackage[dvips]{graphics}
\usepackage{bm}
\usepackage{epsfig}
\usepackage{enumerate}
\usepackage{subfigure}
\usepackage{color}
\usepackage{xcolor}
\usepackage{amsmath,amssymb,amsthm}
\usepackage{sidecap,tikz}
\usepackage{graphicx}
\usepackage{makecell}
\usepackage{hyperref}
\usepackage{multirow}

\usepackage[normalem]{ulem}

\hypersetup{
    pdfnewwindow=true,      
    colorlinks=true,       
    linkcolor=blue,          
    citecolor=magenta,        
    filecolor=blue,      
    urlcolor=blue        
}

\newcommand{\ie}{i.e., }

\newcommand{\bea}{\begin{eqnarray}}
\newcommand{\eea}{\end{eqnarray}}
\newcommand{\beq}{\begin{equation}}  
\newcommand{\eeq}{\end{equation}}

\definecolor{lime}{HTML}{A6CE39}
\DeclareRobustCommand{\orcidicon}{\hspace{-1mm}
	\begin{tikzpicture}
	\draw[lime, fill=lime] (0,0) 
	circle [radius=0.16] 
	node[white] {{\fontfamily{qag}\selectfont \tiny \,ID}};
	\draw[white, fill=white] (-0.0525,0.095) 
	circle [radius=0.007];
	\end{tikzpicture}
	\hspace{-3mm}
}
\foreach \x in {A, ..., Z}{\expandafter\xdef\csname orcid\x\endcsname{\noexpand\href{https://orcid.org/\csname orcidauthor\x\endcsname}
			{\noexpand\orcidicon}}
}

\begin{document} 
\title{ Superconductivity in spin-$3/2$ systems: symmetry classification, odd-frequency pairs, and Bogoliubov Fermi surfaces}

\author{Paramita Dutta\orcidA{}}
\affiliation{Department of Physics and Astronomy, Uppsala University, Box 516, S-751 20 Uppsala, Sweden}
\author{Fariborz Parhizgar\orcidB{}}
\affiliation{Department of Physics and Astronomy, Uppsala University, Box 516, S-751 20 Uppsala, Sweden}
\author{Annica M. Black-Schaffer\orcidC{}}
\affiliation{Department of Physics and Astronomy, Uppsala University, Box 516, S-751 20 Uppsala, Sweden}

\begin{abstract}
The possible symmetries of the superconducting pair amplitude is a consequence of the fermionic nature of the 
Cooper pairs. For spin-$1/2$ systems this leads to the $\mathcal{SPOT}=-1$ classification of superconductivity, 
where $\mathcal{S}$, $\mathcal{P}$, $\mathcal{O}$, and $\mathcal{T}$ refer to the exchange operators for spin, 
parity, orbital, and time between the paired electrons. However, this classification no longer holds for higher spin 
fermions, where each electron also possesses a finite orbital angular momentum strongly coupled with the spin 
degree of freedom, giving instead a conserved total angular moment. For such systems, we here instead introduce 
the $\mathcal{JPT}=-1$ classification, where $\mathcal{J}$ is the exchange operator for the $z$-component of the 
total angular momentum quantum numbers. We then specifically focus on spin-$3/2$ fermion systems and several superconducting cubic half-Heusler compounds that have recently been proposed to be spin-$3/2$ superconductors.
By using a generic Hamiltonian suitable for these compounds we calculate the superconducting pair amplitudes and 
find finite pair amplitudes for all possible symmetries obeying the $\mathcal{JPT}=-1$ classification, including all 
possible odd-frequency (odd-$\omega$) combinations. Moreover, one of the very interesting properties of spin-$3/2$ superconductors is the possibility of them hosting a Bogoliubov Fermi surface (BFS), where the superconducting 
energy gap is closed across a finite area. We show that a spin-$3/2$ superconductor with a pair potential satisfying 
an odd-gap time-reversal product and being non-commuting with the normal-state Hamiltonian hosts both a BFS 
and has finite odd-$\omega$ pair amplitudes. We then reduce the full spin-$3/2$ Hamiltonian to an effective 
two-band model and show that odd-$\omega$ pairing is inevitably present in superconductors with a BFS and vice 
versa.
\end{abstract}

\maketitle

\section{Introduction}
Properties of ordered matter are to a large extent set by the symmetry of the ordered state. For superconductivity, 
this means the symmetry of superconducting Cooper pairs. The symmetry of these Cooper pairs are classified 
following the antisymmetry of the Cooper pair wave function during exchange of all the quantum numbers between 
two constituent electrons, as a result of the fermionic nature of the electrons. This classification of superconducting 
pairs has become known as the $\mathcal{SPOT}=-1$ classification, with $\mathcal{S}$, $\mathcal{P}$, $\mathcal{O}$ and $\mathcal{T}$ referring to the exchange operators for spin, parity, orbital, and time, and is now well established 
in the literature\,\cite{Linder2019,Tanaka2012}. 

Following this classification, Cooper pairs with spin-singlet and spin-triplet spin structures are further identified as 
odd-frequency (odd-$\omega$) or even-frequency (even-$\omega$) pairs, depending on their spatial parity and 
their symmetry under orbital index exchange between the electrons. Odd (even)-$\omega$ superconductivity refers 
to when the Cooper pair amplitude is odd (even) under the exchange of the time coordinates, or equivalently 
frequency, of the two constituent electrons, as introduced by Balatsky and Abrahams~\cite{Balatsky1992} following 
the first prediction of odd-frequency order by Berezinskii in $^3$He~\cite{Berezinskii1974}. Finite odd-$\omega$ pair amplitudes have been predicted to exist mostly in superconducting hybrid structures~\cite{Volkov2003,Tanaka2007_NS,Yokoyama2007,Eschrig2008,Parhizgar2014,Linder2019,Dutta2020a},
 but also in a variety of other systems, such as multiband superconductors~\cite{BlackSchaffer2013,Triola2020,Parhizgar2021}, topological materials~\cite{BlackSchaffer2012,Crepin2015,Cayao2017,Dutta2019,Schmidt2020,Dutta2020b,Parhizgar2020}, 
heavy-fermion compounds~\cite{Coleman1993,Coleman1994,Triola2018}, and driven systems~\cite{Triola2016,Cayao2020}. In several of these systems, experimental consequences of odd-$\omega$ 
pairs have also been explored, including its connection to zero-energy density of states (DOS) peaks~\cite{DiBernardo2015, diBernardo2015_Meissner, Keizer2006,Khaire2010, robinson2010, Krieger2020}.

The Cooper pair symmetry follows the $\mathcal{SPOT}=-1$ classification in all systems where the constituent 
electrons have spin-$1/2$ character. There exist however recent reports on superconductivity in materials where 
the low-energy electrons also carry a finite orbital angular momenta, such that only the total angular momentum 
is a good quantum number\,\cite{Brydon2016,Kim2018}. For example, some cubic half-Heusler compounds, with 
the general form RPtBi or RPdBi where R is the rare earth element, host `spin'-$3/2$ low-energy fermions and 
have been shown to be superconducting at low temperatures~\cite{Brydon2016}. In these materials the conserved 
total angular momentum, or effective, spin-$3/2$ character is protected by high crystal symmetry and spin-orbit 
interaction. These findings have drawn a large amount of attention since the superconducting ground state has 
been found to have several unusual properties, such as a reduced upper critical field~\cite{Kim2018} and a low-temperature penetration depth~\cite{Bay2012} very different from the properties of conventional spin-singlet superconductors~\cite{schrieffer2018theory}. Notably, the spin-$3/2$ character of the constituent electrons of the 
Cooper pairs allows for both exotic spin-quintet with $j=2$ and spin-septet with $j=3$ 
pairing~\cite{Brydon2016, Kim2018, Venderbos2018, Yu2018,Roy2019}, in addition to the spin-singlet ($j=0$) and 
spin-triplet ($j=1$) pairing found in traditional spin-$1/2$ superconductors, where $j$ is the total angular momentum quantum number of the Cooper pair. 

Interestingly, superconductivity in spin-$3/2$ systems can be associated with a Bogoliubov Fermi surface (BFS); 
an inflated spheroidal or toroidal topologically protected region in momentum space across which the energy gap 
is identically zero~\cite{Agterberg2017,Brydon2018,Menke2019,Link2020a}. The reason behind the appearance 
of a BFS has been described in terms of a pseudomagnetic field arising from exotic pairing 
states~\cite{Agterberg2017}. These exotic pairing states have been predicted in a range of materials: nematic superconductors\,\cite{Fu2014}, iron-based superconductors~\cite{Gao2010,Setty2020}, and also cubic half-Heusler compounds~\cite{Tafti2013, nakajima2015, Brydon2016, Xiao2018, Kawakami2018}. The general prescription for 
the appearance of BFS is the existence of an internal electronic degree of freedom, in addition to the spin degree of 
freedom, such as an additional orbital or sub-lattice degree of freedom, subjected to the symmetries of the system~\cite{Brydon2018,Link2020b,Herbut2021}. Higher spin, such as a spin-$3/2$ character, of the low-energy 
bands is perhaps the most explored possibility for that internal degree of freedom to produce a BFS\,\cite{Brydon2018}. 

The BFS manifests itself through a large DOS at zero energy in the superconducting state~\cite{Menke2019}. At 
the same time, zero-energy states have also been shown to be one of the possible signatures of odd-$\omega$ 
pair amplitudes in many systems~\cite{Yokoyama2007,Linder2010,DiBernardo2015}. A natural question then 
arises: is a BFS a manifestation of odd-$\omega$ pairing? This question is also closely related to developing a 
more solid understanding of the exotic pairing present in materials with BFSs. In this work we address these 
issues by focusing on spin-$3/2$ superconductors, and particularly cubic half-Heusler compounds, as they 
present possibilities for both BFS and exotic pairing with higher angular momentum Cooper pairs whose complete symmetry classification is not yet fully developed. 

In particular, we first introduce a complete classification for higher-spin superconductors: $\mathcal{JPT}=-1$, 
where $\mathcal{J}$ is the exchange operator for the quantum numbers associated with the $z$-component of 
the total angular momenta of the constituent electrons of the Cooper pairs. We then establish the applicability of 
this new classification by studying several generic and realistic models describing superconducting half-Heusler compounds and calculating the superconducting pair amplitudes and their symmetries. The $\mathcal{JPT}=-1$ antisymmetry condition gives rise to a total of 32 different classes of Cooper pair symmetry, and we find that 
essentially all can exist in the half-Heusler compounds. In particular, we find all types of possible odd-$\omega$ 
pairs: spin-singlet odd-parity, spin-triplet even-parity, spin-quintet odd-parity, and spin-septet even-parity, which 
in several cases are as large as any even-$\omega$ components. Following this, we derive a general analytical expression for the odd-$\omega$ pair amplitude in spin-$3/2$ systems, which we then use to derive a necessary 
and sufficient condition for finding both odd-$\omega$ pairs and a BFS. Finally, by using a low-energy effective 
two-band model, we show explicitly that the pseudomagnetic field responsible for the BFS also always induces 
odd-$\omega$ pairing and vice versa. This establishes a one-to-one correspondence between BFS and 
odd-$\omega$ pairs in a low-energy model of spin-$3/2$ superconductors.

The rest of this work is organized as follows. We classify the superconducting pair symmetries in Sec.\,\ref{symmetry}
and show the appearance of those symmetries in Sec.~\ref{PA} by considering a generic Hamiltonian suitable for 
half-Heusler compounds (Sec.\,\ref{model}), followed by numerical results for the superconducting pair amplitude (Sec.~\ref{result}). We then find a general analytical expression for odd-$\omega$ pair amplitudes applicable to any 
spin-$3/2$ system in Sec.~\ref{Anal} and explore the connection of the odd-$\omega$ pair amplitudes to BFSs in Sec.\,\ref{BFS}. Finally, we summarize and conclude our results in Sec.\,\ref{conclu}.

\section{Classification of pair symmetry in spin-$3/2$ systems} \label{symmetry}
We begin by developing the classification of the pair symmetry for any superconductor with low-energy electrons of 
spin-$3/2$ character. Central in any such discussion is the superconducting pair amplitude, or Cooper pair amplitude, given by the  anomalous Green's function as 
\bea
\mathcal{F}_{\sigma_{1}a_1;\sigma_{2}a_2}(\boldsymbol{k}_{1},\boldsymbol{k}_{2};t_{1},t_{2})=-i\langle\mathcal{T}_tc_{\sigma_{1}a_1}(\boldsymbol{k}_{1},t_{1}) c_{\sigma_{2}a_2}(\boldsymbol{k}_{2},t_{2})\rangle, \nonumber \\
\label{anoF}
\eea 
where $\mathcal{T}_t$ is the time-ordering operator and $c_{\sigma,a}(\boldsymbol{k},t)$ is the annihilation operator 
for an electron in orbital $a$ with spin $\sigma$ and momentum $\boldsymbol{k}$ at time $t$\,\cite{mahan2013}. Now, 
for regular spin-$1/2$ systems, there exists an antisymmetry condition: 
\beq
\label{eq:SPOT}
\mathcal{SPOT~F}_{\sigma_{1}a_1,\sigma_{2}a_2}(\boldsymbol{k}_{1},\boldsymbol{k}_{2};t_{1},t_{2})=-\mathcal{F}_{\sigma_{1}a_1,\sigma_{2}a_2}(\boldsymbol{k}_{1},\boldsymbol{k}_{2};t_{1},t_{2}),
\eeq 
imposed by the fermionic nature of the electrons of the Cooper pair where the $\mathcal{S}$, $\mathcal{P}$, $\mathcal{O}$, and $\mathcal{T}$ operators are defined as
\bea
\mathcal{S~F}_{\sigma_{1}a_1,\sigma_{2}a_2}(\boldsymbol{k}_{1},\boldsymbol{k}_{2};t_{1},t_{2})
&=&\mathcal{F}_{\sigma_{2}a_1,\sigma_{1}a_2}(\boldsymbol{k}_{1},\boldsymbol{k}_{2};t_{1},t_{2}) \nonumber \\
&=&\pm\mathcal{F}_{\sigma_{1}a_1,\sigma_{2}a_2}(\boldsymbol{k}_{1},\boldsymbol{k}_{2};t_{1},t_{2}), \nonumber \\
\mathcal{P~F}_{\sigma_{1}a_1,\sigma_{2}a_2}(\boldsymbol{k}_{1},\boldsymbol{k}_{2};t_{1},t_{2})
&=&\mathcal{F}_{\sigma_{1}a_1,\sigma_{2}a_2}(\boldsymbol{k}_{2},\boldsymbol{k}_{1};t_{1},t_{2}) \nonumber \\
&=&\pm\mathcal{F}_{\sigma_{1}a_1,\sigma_{2}a_2}(\boldsymbol{k}_{1},\boldsymbol{k}_{2};t_{1},t_{2}), \nonumber \\
\mathcal{O~F}_{\sigma_{1}a_1,\sigma_{2}a_2}(\boldsymbol{k}_{1},\boldsymbol{k}_{2};t_{1},t_{2})
&=&\mathcal{F}_{\sigma_{1}a_2,\sigma_{2}a_1}(\boldsymbol{k}_{1},\boldsymbol{k}_{2};t_{1},t_{2}) \nonumber \\
&=&\pm\mathcal{F}_{\sigma_{1}a_1,\sigma_{2}a_2}(\boldsymbol{k}_{1},\boldsymbol{k}_{2};t_{1},t_{2}), \nonumber \\
\mathcal{T~F}_{\sigma_{1}a_1,\sigma_{2}a_2}(\boldsymbol{k}_{1},\boldsymbol{k}_{2};t_{1},t_{2})
&=&\mathcal{F}_{\sigma_{1}a_1,\sigma_{2}a_2}(\boldsymbol{k}_{1},\boldsymbol{k}_{2};t_{2},t_{1})~~~ \nonumber \\
&=&\pm\mathcal{F}_{\sigma_{1}a_1,\sigma_{2}a_2}(\boldsymbol{k}_{1},\boldsymbol{k}_{2};t_{1},t_{2}).~~~ \nonumber \\
\label{symop}
\eea  
The antisymmetry condition of the pair symmetries can in short form be written as $\mathcal{SPOT=}-1$ and applies 
to any spin-$1/2$ superconductor, where each electron of the Cooper pair possesses zero orbital angular momentum ($l_i=0$) and half spin ($s_i=1/2$) quantum numbers, with $i$ ($=1,2$) being the electron number index. Thus, the 
total angular momentum  for each electron is $j_i=l_i+s_i=1/2$. There are two possible spin-symmetries for spin-$1/2$ systems since $\frac{1}{2} \otimes \frac{1}{2}=0\oplus 1$: spin-singlet with total `spin' $j=s=0$ and $m=0$ and spin-triplet states with total `spin' $j=s=1$ and $m=0,1$. Here $m$ is the quantum number for the $z$-component of the total 
angular momentum, also referred to as secondary total angular momentum quantum number. The Cooper pairs with 
these spin structures can also be identified according to their spatial parities: explicitly, even parity ($s$-wave, 
$d$-wave etc.) and odd parity ($p$-wave, $f$-wave etc.) pairs. Given the spin structure and spatial parity, the Cooper 
pair symmetry can further be classified according to its evenness or oddness under the orbital index (if such exists) and 
then finally with respect to the time/frequency dependence. We note that pair amplitudes that are odd with respect to 
the change of the time (or equivalently frequency) of the two electrons are identified as odd-$\omega$ pair amplitudes. 
As an example, all spin-triplet even-parity and all spin-singlet odd-parity  states are necessarily odd-$\omega$ states 
when they are even with respect to  exchange of the orbital index\,\cite{Linder2019,BlackSchaffer2013}. Similarly, we 
can also find other possible odd-$\omega$ pair symmetries when the pair amplitude is odd in orbital index\,\cite{BlackSchaffer2013}.

With the above outline of the pair symmetry classification for spin-$1/2$ fermions, we now move on to the generalization 
of the antisymmetry condition for higher spin systems, since the condition Eq.\,\eqref{eq:SPOT} no longer remains applicable when the spin and orbital are strongly coupled to each other. In fact, for such strong spin-orbit coupled systems, both the spin and orbital angular momentum are not good quantum numbers at all. Instead, the total angular momentum 
of each electron is a good quantum number, and it becomes natural to consider that instead when identifying the superconducting pair symmetry. More precisely, when two electrons of total angular momenta $j_1$ and $j_2$ pair, the total angular momentum quantum number $j$ for the paired state is constrained by the condition $|j_1-j_2|\le j\le (j_1+j_2)$. The quantum number associated with the $z$-component of the total angular momentum $m$ follows $
-j \le m \le j$. It also satisfies $m=m_1+m_2$ where $m_1$ and $m_2$ are the quantum numbers for the $z$-component of the total angular momentum of the two electrons, individually satisfying $-j_i\le m_i \le j_i$. The paired state $|j,m\rangle$ can be expanded in the basis $|j_1,j_2;m_1,m_2\rangle$ using the completeness and normalization 
conditions as
\bea
|j,m\rangle&=& \sum\limits_{m_1=-j_1}^{j_1}  \sum\limits_{m_2=-j_2}^{j_2} |j_1,j_2;m_1,m_2\rangle \langle j_1,j_2;m_1,m_2| j,m\rangle \nonumber \\ 
&=&\sum\limits_{m_1,m_2} \underbrace{ \langle j_1,j_2;m_1,m_2 |j,m\rangle}_{\text {C.G. coefficient}} |j_1,j_2;m_1,m_2\rangle 
\label{cg}
\eea
where the Clebsch-Gordon (C.G.) coefficients are scalar numbers~\cite{griffithsQM}. 

In particular, we are here interested in strongly spin-orbit coupled systems where the low-energy band structure has 
a spin-$3/2$ character~\cite{Savary2017,Kim2018}. When two such electrons with $j_1=3/2$ and $j_2=3/2$ couple 
to form a Cooper pair, the possible states can be found following the product $\frac{3}{2} \otimes \frac{3}{2}=0\oplus 1\oplus 2 \oplus 3$~\cite{Brydon2016,Yang2016,Savary2017,Kim2018,Venderbos2018,Yu2018}. Thus, for such 
higher spin system, the pair symmetries are enriched by five spin-quintet ($j=2$) and seven spin-septet ($j=3$) states, 
in addition to extended spin-singlet ($j=0$) and extended spin-triplet ($j=1$) states.\footnote{One can here raise the question about the possibility of one electron having $j_1=1/2$ and another having $j_2=3/2$ character. We avoid considering this case as it is unlikely a Fermi surface includes two electrons of such different spin characters.} 
Henceforth, we use the short notation $|m_1,m_2\rangle$ instead of $|j_1,j_2;m_1,m_2\rangle$ for brevity since $j_1=j_2=3/2$.

With the concept of the spin structures of the Cooper pairs for spin-$3/2$ fermions clear, the next natural question is about the overall symmetry of the pairing. We note that, for spin-$1/2$ systems, we have four symmetry operators $\mathcal{S}$, $\mathcal{P}$, $\mathcal{O}$, and $\mathcal{T}$ corresponding to the spin quantum number, spatial parity, orbital, and time for the Cooper pairs. However, due to the strong spin-orbit coupling in the spin-$3/2$ systems, we can now not separately identify the spin and orbital quantum numbers. Instead, the total angular momentum and its $z$-component 
are the only good quantum numbers. Thus, we now rewrite the anomalous Green's function of the Cooper pair of Eq.\eqref{anoF} characterized by quantum numbers for angular momentum, spatial parity, and time as
\beq
\mathcal{F}_{j_{1}m_1,j_{2}m_2}(\boldsymbol{k}_{1},\boldsymbol{k}_{2};t_{1},t_{2})=-i\langle\mathcal{T}_t c_{j_{1}m_1}(\boldsymbol{k}_{1},t_{1}) c_{j_{2}m_2}(\boldsymbol{k}_{2},t_{2})\rangle
\label{anoFnew}
\eeq
where $c_{jm}(\boldsymbol{k},t)$ is now the annihilation operator for an electron with the total angular momentum quantum numbers $j$, $m$ and spatial momentum $\boldsymbol{k}$ at time $t$. Note that each total angular momentum quantum number $j_i$ allows various states identified by $m_i$ following Eq.\,\eqref{cg} and thus we must include $m_i$ index too. We then introduce a symmetry operator for the exchange of the total angular momentum quantum numbers of the two electrons, $\mathcal{J}$, which effectively exchanges only the $z$-components of the total angular momentum quantum number, since here $j_1=j_2=3/2$. Thus we can now have Cooper pairs which are even or odd with respect to the exchange of the 
quantum numbers $m_1\leftrightarrow m_2$ as
\bea
\mathcal{JF}_{j_{1} m_1,j_{2}m_2}(\boldsymbol{k}_{1},\boldsymbol{k}_{2};t_{1},t_{2})
&=&\mathcal{F}_{\frac{3}{2} m_2,\frac{3}{2} m_1}(\boldsymbol{k}_{1},\boldsymbol{k}_{2};t_{1},t_{2}) \nonumber \\
&=&\pm\mathcal{F}_{\frac{3}{2} m_1,\frac{3}{2} m_2}(\boldsymbol{k}_{1},\boldsymbol{k}_{2};t_{1},t_{2}). \nonumber \\
\label{jop}
\eea 
On top of this, the pair amplitudes can again be even or odd with respect to the spatial parity $\mathcal{P}$, as well as 
(relative) time exchange $\mathcal{T}$ operations ,exactly similar to spin-$1/2$ fermions, see Eq.\eqref{symop}. We note that, 
there is now no separate symmetry operation for the orbital index since it is no longer a good quantum 
number here, but the orbital index is instead effectively included into the $\mathcal{J}$ operation. Finally, 
all these exchange operations should follow an antisymmetry condition maintaining the fermionic property of the 
electrons as
\bea
\mathcal{JPT\,F}_{\frac{3}{2}m_1, \frac{3}{2} m_2}(\boldsymbol{k}_{1},\boldsymbol{k}_{2};t_{1},t_{2})=-\mathcal{F}_{\frac{3}{2}m_1,\frac{3}{2} m_2}(\boldsymbol{k}_{1},\boldsymbol{k}_{2};t_{1},t_{2}). \nonumber \\
\label{jpt}
\eea
In short this complies with $\mathcal{JPT}=-1$, which identify the evenness or oddness of all the possible pairing for spin-$3/2$ systems with respect to the angular momentum $\mathcal{J}$, spatial parity $\mathcal{P}$ and time $\mathcal{T}$ or frequency. 

Having derived the $\mathcal{JPT}=-1$ condition, we next illustrate how to classify the Cooper pair states following this 
explicit antisymmetry condition. For this, we first explicitly write out the possible states by 
using Eq.\,\eqref{cg}, following Ref.~[\onlinecite{Kim2018}], and then show their classification in terms of $\mathcal{JPT}=-1$ in Table\,\ref{CPdetails}.
\begin{table*}[ht]
\begin{tabular}{|c|c|c|c|c|c|c|}
   \hline
   \hline
 Class &  Pairing state & Cooper Pair & Angular momentum & Parity   & Freq./Time \\
 &   & $|j_1,m_1;j_2,m_2\rangle\equiv |m_1,m_2\rangle$ & ($\mathcal{J}$) & ($\mathcal{P}$)   & ($\mathcal{T}$) \\
   \hline \hline
1. &  Singlet (sing.) &  &   &Even& Even \\
&($j=0$; $m$=0)& $\tfrac{1}{2}\left (|\tfrac{3}{2},-\tfrac{3}{2}\rangle-|-\tfrac{3}{2},\tfrac{3}{2}\rangle-|\tfrac{1}{2},-\tfrac{1}{2}\rangle+|-\tfrac{1}{2},\tfrac{1}{2}\rangle \right )$  &  Odd &------------ &------------------ \\
&  &  &   & Odd & Odd \\
 \hline  \hline
  2. &   Triplet\,($t_1$) &&&Even&   Odd\\
& ($j=1$; $m$=-1)& $ \frac{1}{\sqrt{10}}\left (\sqrt{3}|-\tfrac{3}{2},\tfrac{1}{2}\rangle-2|-\tfrac{1}{2},-\tfrac{1}{2}\rangle+\sqrt{3}|\tfrac{1}{2},-\tfrac{3}{2}\rangle\right )$ &Even &------------ &------------------\\ 
 &  &  &   & Odd & Even \\
\hline
 3. &  Triplet\,($t_2$) & & &Even &  Odd \\
& ($j=1$; $m$=0)& $ \frac{1}{\sqrt{20}}\left (3|\tfrac{3}{2},-\tfrac{3}{2}\rangle-|\tfrac{1}{2},-\tfrac{1}{2}\rangle-|-\tfrac{1}{2},\tfrac{1}{2}\rangle+3|-\tfrac{3}{2},\tfrac{3}{2}\rangle \right)$ &Even&------------ &------------------\\ 
&  &  &   & Odd & Even \\
   \hline
4. &   Triplet\,($t_3$) &  &&Even &  Odd \\
& ($j=1$; $m$= 1)& $\frac{1}{\sqrt{10}}\left (\sqrt{3}|\tfrac{3}{2},-\tfrac{1}{2}\rangle-2|\tfrac{1}{2},\tfrac{1}{2}\rangle+\sqrt{3}|-\tfrac{1}{2},\tfrac{3}{2}\rangle\right ) $ &Even&------------ &------------------\\ 
&  &  &   & Odd &  Even \\
   \hline \hline
5. & Quintet\,($q_1$) && &Even&Even\\
&  ($j=2$; $m$=-2)&$\frac{1}{\sqrt{2}}\left (|- \tfrac{3}{2},-\tfrac{1}{2}\rangle-|-\tfrac{1}{2},- \tfrac{3}{2}\rangle \right )$&Odd&------------ &------------------\\
&  &  &   & Odd &  Odd \\
\hline
6. & Quintet\,($q_2$) & &&Even& Even \\
 & ($j=2$; $m$=-1)&$\frac{1}{\sqrt{2}}\left (|- \frac{3}{2},\frac{1}{2}\rangle-|\frac{1}{2},- \frac{3}{2}\rangle \right)$&Odd&------------ &------------------\\
&  &  &   & Odd & Odd \\
\hline
7. & Quintet\,($q_3$) && &Even& Even \\
 &($j=2$; $m$=0)&$ \frac{1}{2}\left (| \tfrac{3}{2},- \tfrac{3}{2}\rangle+|\tfrac{1}{2},-\tfrac{1}{2}\rangle-|-\tfrac{1}{2},\tfrac{1}{2}\rangle-|- \tfrac{3}{2}, \tfrac{3}{2}\rangle \right) $&Odd &------------ &------------------\\
&  &  &   & Odd &  Odd  \\
 \hline
8. & Quintet\,($q_4$) & &&Even& Even \\
 & ($j=2$; $m$=1)&$ \frac{1}{\sqrt{2}}\left (| \tfrac{3}{2},-\tfrac{1}{2}\rangle-|-\tfrac{1}{2}, \tfrac{3}{2}\rangle \right )$&Odd&------------ &------------------\\
&   &  &   &.Odd & Odd \\
 \hline
9. & Quintet\,($q_5$) & &&Even& Even \\
& ($j=2$; $m$=2)&$\frac{1}{\sqrt{2}}\left (| \tfrac{3}{2},\tfrac{1}{2}\rangle-|\tfrac{1}{2}, \tfrac{3}{2}\rangle \right ) $&Odd&--------------- &------------------\\
 & &  &   & Odd & Odd \\
 \hline \hline
10. &  Septet\,($s_1$) & &&Even& Odd\\
&  ($j=3$; $m$=-3)&$ |-\tfrac{3}{2},-\tfrac{3}{2}\rangle$&Even&------------ &------------------\\
 &  &  &   & Odd & Even \\
 \hline
11. &  Septet\,($s_2$) &&&Even&  Odd\\
&  ($j=3$; $m$=-2)&$\ \frac{1}{\sqrt{2}}\left (|-\tfrac{3}{2},-\tfrac{1}{2}\rangle+|-\tfrac{1}{2},-\tfrac{3}{2}\rangle \right  ) $&Even&------------ &------------------\\
 &  &  &   & Odd & Even  \\
 \hline
12. &  Septet\,($s_3$) & &&Even&  Odd \\
&  ($j=3$; $m$=-1)&$ \frac{1}{\sqrt{5}}\left (|-\tfrac{3}{2},\tfrac{1}{2}\rangle+\sqrt{3}|-\tfrac{1}{2},-\tfrac{1}{2}\rangle+|\tfrac{1}{2},-\tfrac{3}{2}\rangle\right )$ &Even&------------ &------------------\\
&   &  &   &Odd &  Even \\
 \hline
13. &  Septet\,($s_4$) & &&Even& Odd \\
 &   ($j=3$; $m$=0)&$ \frac{1}{\sqrt{20}}\left (|\tfrac{3}{2},-\tfrac{3}{2}\rangle+3|\tfrac{1}{2},-\tfrac{1}{2}\rangle+3|-\tfrac{1}{2},\tfrac{1}{2}\rangle+|-\tfrac{3}{2},\tfrac{3}{2}\rangle \right) $ &Even&------------ &------------------\\
&     &  &   & Odd &  Even \\
 \hline
14. & Septet\,($s_5$) & &&Even&  Odd \\
 &  ($j=3$; $m$=1)&$\frac{1}{\sqrt{5}}\left (|\tfrac{3}{2},-\tfrac{1}{2}\rangle+\sqrt{3}|\tfrac{1}{2},\tfrac{1}{2}\rangle+|-\tfrac{1}{2},\tfrac{3}{2}\rangle\right )$ &Even&------------ &------------------\\
&   &  &   & Odd & Even \\
 \hline
15. &  Septet\,($s_6$) & &&Even&  Odd \\
 &  ($j=3$; $m$=2)&$ \frac{1}{\sqrt{2}}(|\frac{3}{2},\frac{1}{2}\rangle+|\frac{1}{2},\frac{3}{2}\rangle)$ &Even&------------ &------------------\\
&   &  &   & Odd & Even \\
 \hline
16. &  Septet\,($s_7$) & &&Even & Odd \\
&   ($j=3$; $m$=3)&$ |\frac{3}{2},\frac{3}{2} \rangle$ &Even&------------ &------------------\\
 &   &  &   & Odd & Even \\
 \hline \hline
 \end{tabular}
 \caption{Complete symmetry classification of Cooper pairs for $j=3/2$ fermions following the antisymmetry condition 
 $\mathcal{JPT}=-1$. Cooper pair configurations in column $3$ follow Ref.~[\onlinecite{Kim2018}].}
\label{CPdetails}
\end{table*}
Note that, while the pairing state $|m_1,m_2\rangle$ may be either a single state or a combination of two states or 
more, as seen in the third column in Table\,\ref{CPdetails}, the operation of $\mathcal{J}$ ($m_1\leftrightarrow m_2$) 
is additive for each state and thus it always gives either an even or odd Cooper pair state with respect to this exchange. Overall, in Table\,\ref{CPdetails} there are one singlet, three triplet ($t_1$, $t_2$, $t_3$), five quintet ($q_1$, $q_2$, $q_3$, $q_4$, $q_5$), and seven septet ($s_1$, $s_2$, $s_3$, $s_4$, $s_5$, $s_6$, $s_7$) states, such that a total 
of $16$ possible spin structures exist for spin-$3/2$ fermions. Considering also spatial parity and time, each of these 
$16$ states can be further characterized either by being odd or even under parity and time, resulting in a maximum of $32$ possible states for spin-$3/2$ fermions, as shown in Table\,\ref{CPdetails}.

We next explicitly illustrate some of the symmetries of Table\,\ref{CPdetails}. The spin-singlet state for spin-$3/2$ 
fermions can be thought of as a combination of two separate states formed by the first two and last two parts of the Cooper pair as: 
$\frac{1}{2}\left (|\tfrac{3}{2},-\tfrac{3}{2}\rangle-|-\tfrac{3}{2},\tfrac{3}{2}\rangle\right)$ and $\frac{1}{2}\left(-|\tfrac{1}{2},-\tfrac{1}{2}\rangle+|-\tfrac{1}{2},\tfrac{1}{2}\rangle \right)$, also called as an extended singlet state\,\cite{Kim2018}. 
When the $\mathcal{J}$ operator operates on each of these combined states, it exchanges the $m_i$ quantum 
numbers of the two electrons of each individual parts following Eq.\eqref{jop} \ie $3/2 \leftrightarrow -3/2$ for 
the first state and $1/2 \leftrightarrow -1/2$ for the last state and both combined states are thus odd with respect to 
this exchange operation. Since the operation of $\mathcal{J}$ is additive, the whole singlet state is also odd. Now, 
this odd state can further be both even and odd with respect to the spatial parity $\mathcal{P}$ and in time 
$\mathcal{T}$ or frequency. Following $\mathcal{JPT}=-1$, the singlet state has to be either even-parity 
even-$\omega$ or odd-parity odd-$\omega$, as also seen in Table.\,\ref{CPdetails}.

Next we consider one of the triplet states, triplet\,$t_1$, in Table\,\ref{CPdetails}, which consists of two parts: 
$\frac{\sqrt{3}}{10}(|-\tfrac{3}{2},\tfrac{1}{2}\rangle+|\tfrac{1}{2},-\tfrac{3}{2}\rangle)$ and $-\frac{2}{10}|-\tfrac{1}{2},-\tfrac{1}{2}\rangle$. When $\mathcal{J}$ operates on these combined states, the corresponding exchange operations 
are $-3/2 \leftrightarrow 1/2$ and $-1/2 \leftrightarrow -1/2$, in the first and last states respectively, and these 
exchange operations show that the pair amplitude for this triplet state $t_1$ is even with respect to $\mathcal{J}$. 
Note that, in the latter case, the state is self-exchanging under the operation $-1/2 \leftrightarrow -1/2$, since both 
the electrons are characterized by the same $z$-components of the total angular momentum quantum numbers. 
The other two spin-triplets, triplet\,$t_2$ and $t_3$, are also even states under the exchange operation $\mathcal{J}$ 
in the same way. Then, according to the $\mathcal{JPT}=-1$ antisymmetry condition, all the spin-triplet states can be either even-parity odd-$\omega$ or odd-parity even-$\omega$ states. Similarly, we classify all the spin-quintet ($q_j$) 
and spin-septet ($s_j$) states. All the five spin-quintet states are odd in $\mathcal{J}$. These spin-quintet states can 
thus further be categorized as either even-parity even-$\omega$ or odd-parity odd-$\omega$, similar to the spin-singlet state. Similarly, all the seven spin-septets can be either even-parity odd-$\omega$ or odd-parity even-$\omega$. Note that, septet\,$s_1$ and $s_7$ states are states that change into themselves under the exchange operation by $\mathcal{J}$, thus always resulting  in $\mathcal{J}=+1$. All of these symmetries are explicitly written out in Table\,\ref{CPdetails}.

\section{Examples of pair symmetries for spin-$3/2$ fermions} \label{PA}
With the classification for spin-$3/2$ fermion systems in the previous section, we now show the existence of all 
those pair symmetries by considering a generic normal-state Hamiltonian suitable for half-Heusler compounds 
which are known to both host low-energy spin-$3/2$ fermions due to the strong spin-orbit coupling and be superconducting~\cite{Brydon2016}. We here consider several different superconducting pair potentials following 
the superconducting behaviors of these materials\,\cite{Brydon2016,Brydon2018}.
\subsection{Model for superconducting spin-$3/2$ fermion systems} \label{model}
We start with the Bogoliubov de-Gennes (BdG) Hamiltonian given by
\bea
\check{H}=\frac{1}{2} \sum\limits_{\boldsymbol k}  \Psi_{\boldsymbol k}^{\dagger} \check{\mathcal{H}}_{\boldsymbol k}\Psi_{\boldsymbol k}
\eea
where $\Psi_{\boldsymbol k}=(c_{\boldsymbol k},c_{-{\boldsymbol k}}^{\dagger})^T$ is the Nambu spinor with $c_{\boldsymbol k}$ being a four component spinor encoding the internal degrees of freedom for spin-$3/2$ 
fermions, $c _{\boldsymbol k}=(c_{{\boldsymbol k},3/2}, c_{{\boldsymbol k},1/2}, c_{{\boldsymbol k},-1/2}, c_{{\boldsymbol k},-3/2})^T$, and $\boldsymbol{k}=\{k_x,k_y,k_z\}$. Here
\bea
\check{\mathcal{H}}_{\boldsymbol{k}}=\begin{pmatrix}
\hat{H}_0(\boldsymbol{k})& \hat{\Delta}(\boldsymbol{k})\\
\hat{\Delta}^{\dagger}(\boldsymbol{k})&-\hat{H}^T_0(\boldsymbol{-k})
\end{pmatrix}.
 \label{eq:bdgH}
\eea
where we label $8\times 8$ and $4\times 4$ matrices operator by $\check{...}$ and $\hat{...}$, respectively. We  here consider cubic materials where the low-energy bands of a strongly spin-orbit coupled system can be described by a generic $\boldsymbol{k.p}$ model Hamiltonian appropriate for half-Heusler materials, but neglecting higher order 
terms as they do not affect our conclusions qualitatively\,\cite{Luttinger1955,Agterberg2017}. In particular, we write 
the normal part of the BdG Hamiltonian\,\cite{Agterberg2017} as
\bea
\hat{H}_0(\boldsymbol{k})&=&\alpha k^2 \hat{I}+\beta \sum\limits_{\nu} k_{\nu}^2 \hat{J}_{\nu}^2+\gamma \sum\limits_{\nu \neq \nu^{\prime}} k_{\nu} k_{\nu^{\prime}} \hat{J}_{\nu} \hat{J}_{\nu^{\prime}}\nonumber\\
&+& \delta \sum\limits_{\nu} k_{\nu} (\hat{J}_{\nu+1} \hat{J}_{\nu} \hat{J}_{\nu+1}-\hat{J}_{\nu+2} \hat{J}_{\nu} \hat{J}_{\nu+2}) -\mu,~~~~~
\label{ham_N}
\eea
where $\hat{I}$ is the identity operator and $\hat{J}_{\nu}$, with $\nu\in \{x, y, z\}$, are the $4 \times 4$ matrices for 
spin-$3/2$ fermions (see Appendix\,\ref{j_operators}). Here $\nu + 1$ and $\nu+2$ follow the cyclic shift operations 
on $\{x, y, z\}$. For each momentum $\boldsymbol{k}$, $\hat{H}_0(\boldsymbol{k})$ thus has four electronic degrees 
of freedom, arising from the four channels with $m=\pm 3/2$ and $\pm 1/2$. Here, $\alpha$, $\beta$, $\gamma$, and $\delta$ are all real constants and material dependent. The kinetic energy part involves $\alpha$, while $\beta$ and $\gamma$ characterize the symmetric spin-orbit coupling strength. Moreover, $\delta$ is proportional to the asymmetric 
part of the spin-orbit interaction and it breaks the inversion symmetry of the normal state Hamiltonian and (as we will 
see) play a vital role for generating an odd-parity pair amplitude. We here choose the parameter values: $a=1$, 
$\alpha=20.5(a/\pi)^2$eV, $\beta=-18.5 (a/\pi)^2$\,eV, $\gamma=-12.7 (a/\pi)^2$\,eV, and $\delta=0.06 (a/\pi)$\,eV as 
they nicely capture all the essential features of the band structures for YPtBi, which is one of the superconducting 
materials showing spin-$3/2$ character~\cite{Brydon2016}. We further fix the chemical potential $\mu=50$\,meV 
(intrinsic doping) as it encodes the spin-orbit split hole-like Fermi surface consistent with literature\,\cite{Kim2018}. 
We note that our results of the pair symmetry classification are robust to changes in the parameter values and the behavior of 
the pair amplitudes also remain qualitatively similar.

Having the Hamiltonian in Eq.~\eqref{eq:bdgH} we define the Green's function as
\begin{align}
\check{G} = (i \omega - \check{H})^{-1} = \begin{pmatrix}
{\cal \hat{G}}&{\cal \hat{F}}\\
\hat{ \bar{\mathcal{F}}}&\hat{ \bar{\mathcal{G}}}
\end{pmatrix}
\label{eq:Green}
\end{align}
where $\mathcal{\hat{G}}$ ($\mathcal{\hat{\bar{G}}}$) and $\mathcal{\hat{F}}$ ($\mathcal{\hat{\bar{F}}}$) are the 
$4\times 4$ normal and anomalous Green's functions in particle (hole) space, respectively. For superconductivity 
we are particularly interested in $\mathcal{F}$, where each element are characterized by ${j_1,m_1;j_2,m_2}$ as 
explicitly written in Eq.\,\eqref{anoFnew}. 

\subsection{Numerical results for pair amplitudes} \label{result}
To establish the existence of the symmetry classes in Table\,\ref{CPdetails}, we consider different superconducting 
pair potentials of $\hat{\Delta}(\boldsymbol{k})$ for the BdG Hamiltonian Eq.\,\eqref{eq:bdgH} and calculate the 
anomalous Green's function using Eq.\,\eqref{eq:Green}. We start with the simplest even-parity $s$-wave pair 
potential and then also consider odd-parity $p$-wave, followed by chiral even-parity ($d$-wave) pair potentials 
since all these have been suggested to describe spin-$3/2$ superconductors in the current literature~\cite{Brydon2016,Brydon2018,Kim2018,Boettcher2016}.

\subsubsection{Even-parity ($s$-wave) pair potential }\label{sec:swave}
We start by considering a momentum independent $s$-wave superconducting pair potential as it is the simplest 
possible form. For spin-$3/2$ systems it takes the form\,\cite{Brydon2018}
\begin{align}
\hat{\Delta}(\boldsymbol{k})=\Delta_s U_T
\label{eq:swave}
\end{align}
with the $s$-wave gap $\Delta_s$ being a real constant and the unitary operator $U_T$ defined in Appendix~\ref{j_operators}. 

\begin{figure}[!thpb]
\centering
\includegraphics[scale=0.695]{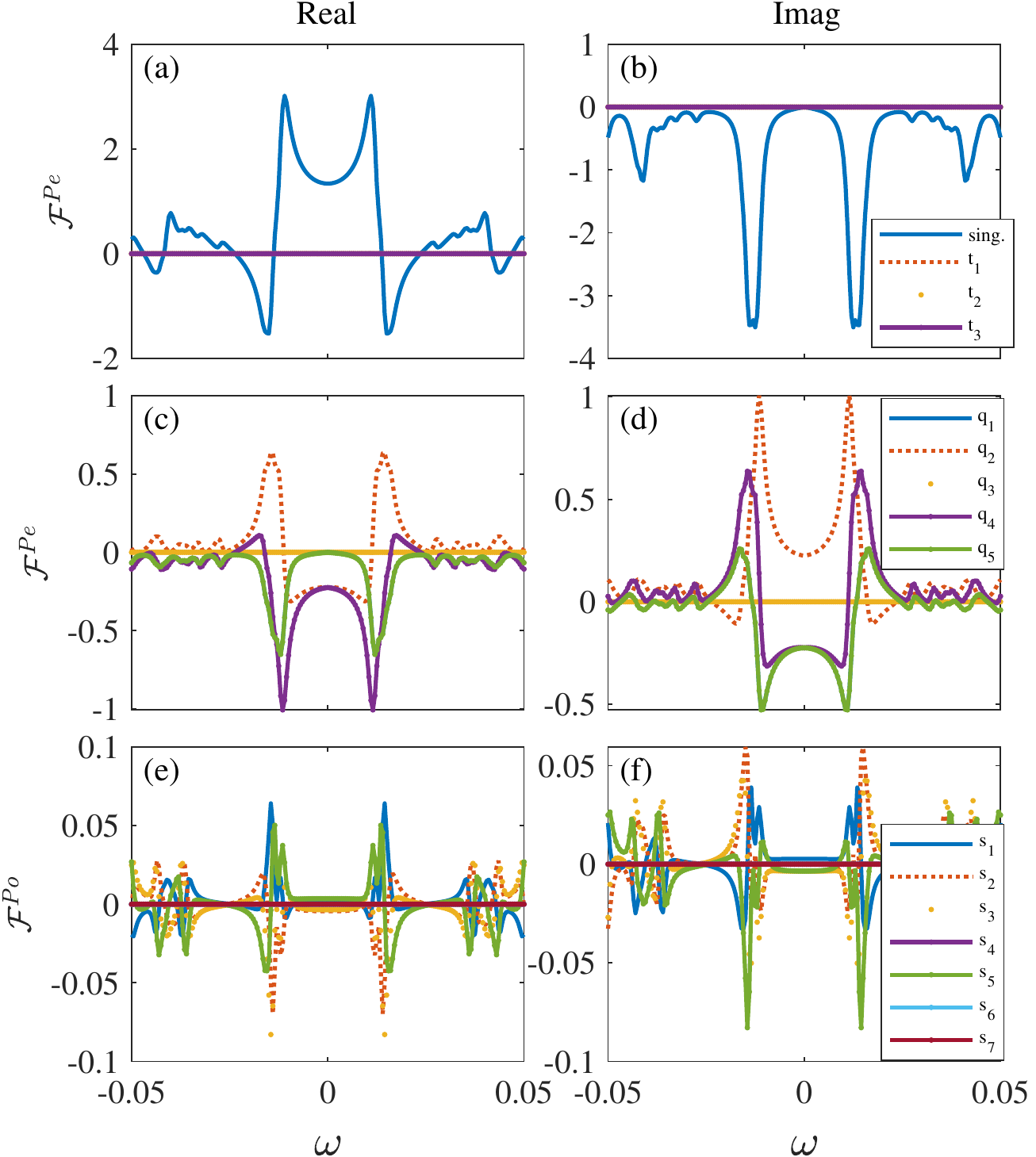}
\caption{Real (a,c,e) and imaginary (b,d,f) part of $\mathcal{F}^{Pe}$ (a-d) and $\mathcal{F}^{Po}$ (e,f) as a function 
of frequency $\omega$ considering the pair potential of Eq.\,\eqref{eq:swave} with $\Delta_s=0.01$eV. The rest of the pair amplitudes are zero.}
\label{fig:swave}
\end{figure}

We calculate the real and imaginary parts of the anomalous Green's function following Eq.\,\eqref{eq:Green} in order to 
extract the pair amplitudes and present them in Fig.\,\ref{fig:swave}. To avoid any mixing of spatial parity, we 
always symmetrize $\mathcal{\hat{F}}$ by breaking it up into its even- and odd-parity parts: 
$\mathcal{F}^{Pe} = \mathcal{F}(\boldsymbol{k})+\mathcal{F}(\boldsymbol{-k})$ and $\mathcal{F}^{Po}=\mathcal{F}(\boldsymbol{k})-\mathcal{F}(\boldsymbol{-k})$. We here take the summation over positive $\boldsymbol{k}$ from zero to a cut-off value of the momentum $k_c=\pi/2$, but our results do not qualitatively depend on the cut-off. From both the real 
and imaginary parts of the pair amplitude in Figs.~\ref{fig:swave}\,(a-b) we see that the spin-singlet even-parity pair amplitude has an even frequency dependency (even-$\omega$) with the higher values of the pair 
amplitude appearing for lower values of $\omega$. On the other hand, all the spin-triplet even-parity pair amplitudes are zero. For all remaining the figures in this section, if a pair amplitude is not plotted, it is because it is identically zero. 

Similar to the spin-singlet pair amplitude, all five spin-quintet pair amplitudes in Figs.\,\ref{fig:swave}\,(c-d) are also even-$\omega$, since they are even in parity. Note that one of the spin-quintets, quintet\,$q_3$, is zero, while the $q_1$ and $q_5$ pair amplitudes are equal to each other. Similar to the spin-singlet amplitudes, three of the spin-quintet pair amplitudes also show a large peak for low $\omega$.
Finally, moving on to the spin-septet part in Fig\,\ref{fig:swave}, we find that all the spin-septet even-parity pair amplitudes are identically zero. Instead, it is the spin-septet odd-parity pair amplitudes that exist for a $s$-wave 
pair potential. Note that three of the spin-septets, $s_4$, $s_6$, and $s_7$, are here identically zero. However, a basis rotation will interchange which spin-septets are zero, and we can in this way confirm that all spin-septet amplitudes are odd-parity and even-$\omega$. A similar procedure and result are also present for the zero spin-quintet amplitude discussed above.
\begin{figure}[!thpb]
\centering
\includegraphics[scale=0.695]{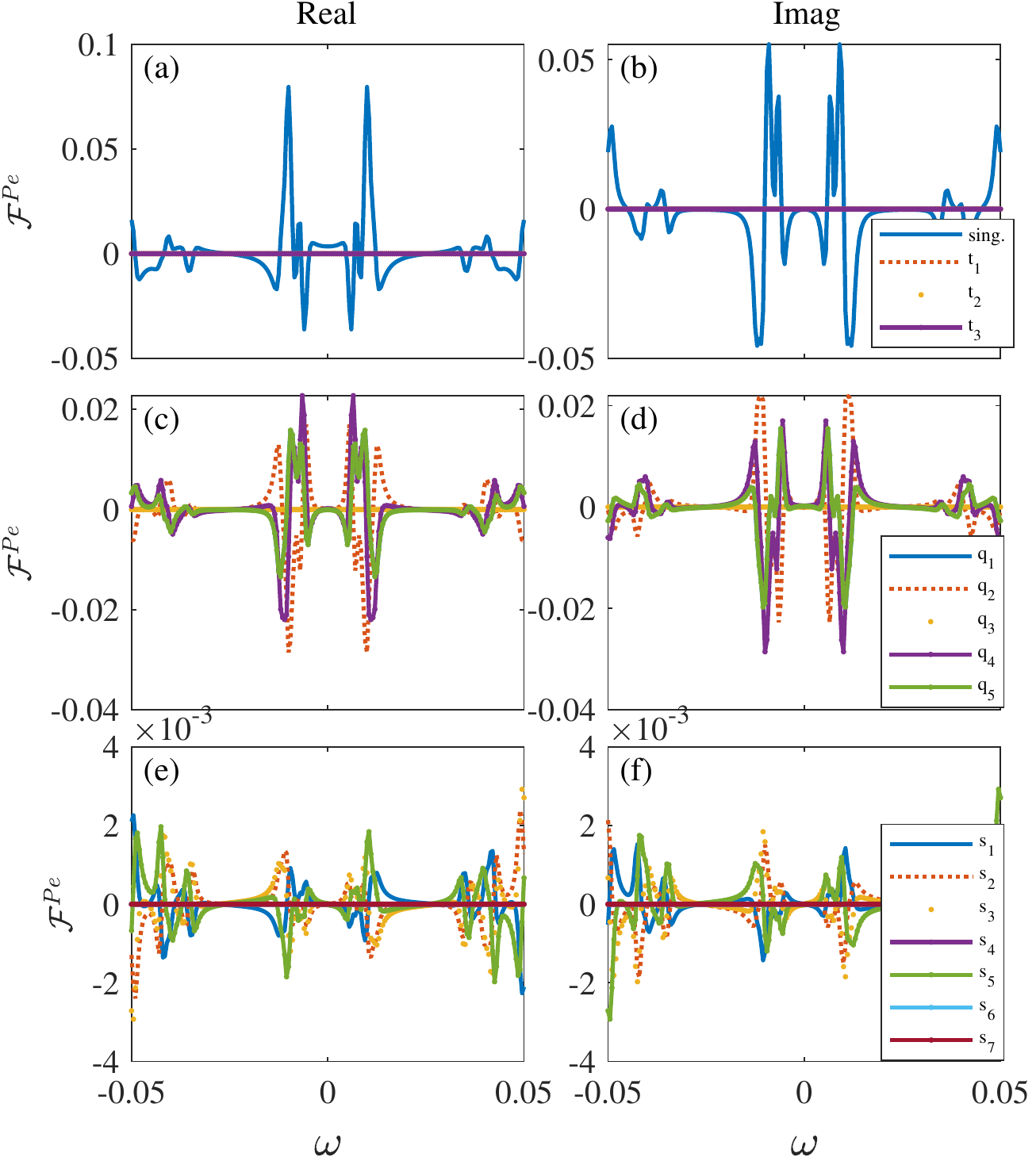}
\caption{Real (a,c,e) and imaginary (b,d,f) part of $\mathcal{F}^{Pe}$ as a function of frequency $\omega$ using the pair potential of Eq.\,\eqref{eq:pwave} with $\Delta_p=0.01$eV. }
\label{fig:pwave+}
\end{figure}

Based on the results in Fig.\,\ref{fig:swave} we conclude that the spin-singlet even-parity pair amplitude is 
even-$\omega$ and very similar to what happens in spin-$1/2$ systems\,\cite{Linder2019}. Additionally, we 
find spin-quintet even-parity even-$\omega$ pair amplitudes and spin-septet odd-parity even-$\omega$ pair 
amplitudes, which are both unique for higher spin systems. The non-zero spin-quintet and spin-septet pair 
amplitudes are similar in magnitude but smaller than that of spin-singlet pair amplitude. All these pair symmetries 
found for the $s$-wave pair potential match with the symmetries shown in Table\,\ref{CPdetails}, explicitly, the 
even-parity parts of classes:\,1, 5-6, 8-9 and odd-parity parts of classes:\,10-12, 14. 
Notably, there is no odd-$\omega$ pair amplitude (irrespective of parity) present for the $s$-wave pair potential. We explain this absence in the Section \ref{Anal}.

\subsubsection{Odd-parity ($p$-wave) pair potential }
\begin{figure}[!thpb]
\centering
\includegraphics[scale=0.695]{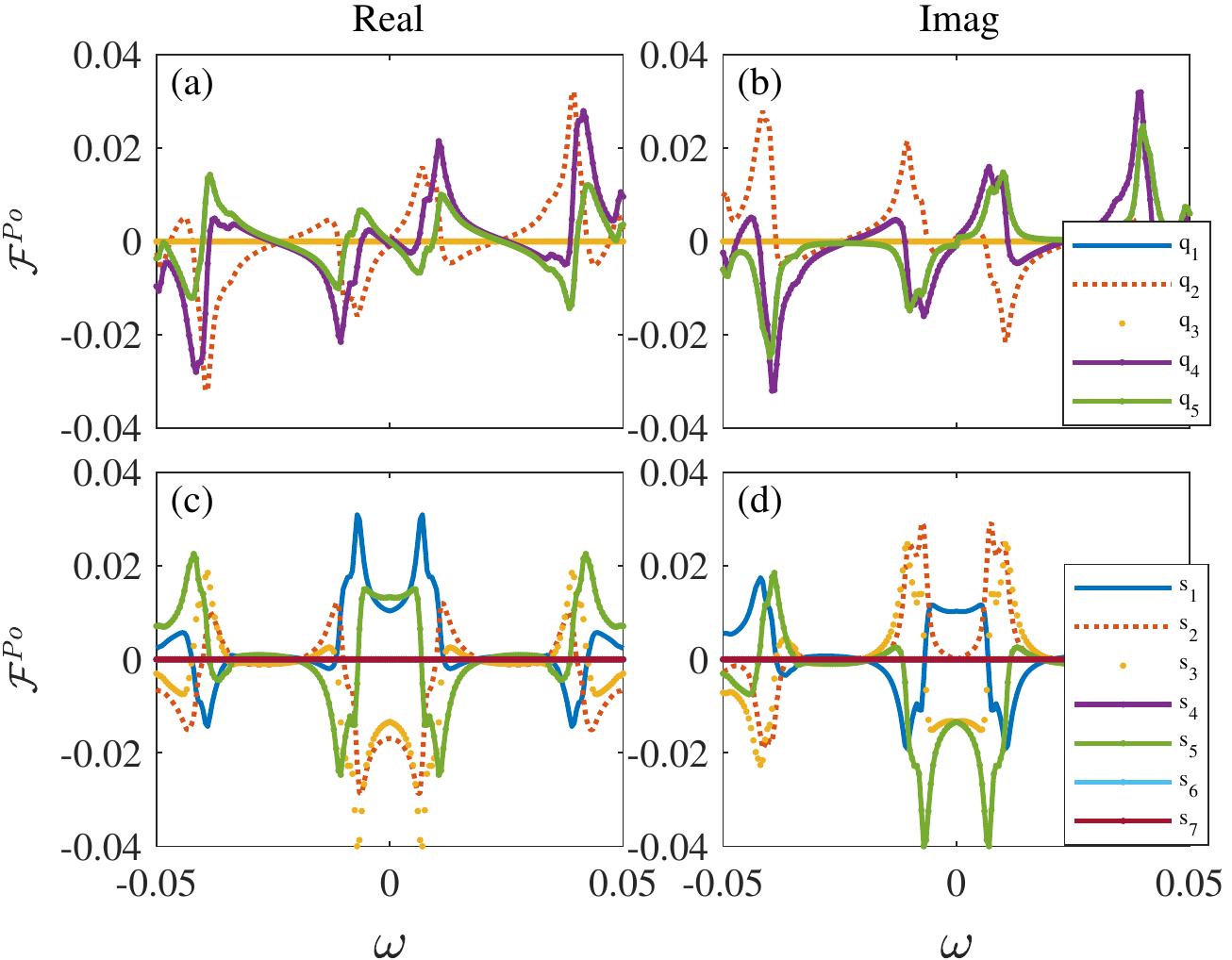}
\caption{Real (a,c) and imaginary (b,d) part of $\mathcal{F}^{Po}$ as a function of frequency $\omega$. The rest of 
the odd-parity pair amplitudes are zero. The parameter values are same as in Fig.\,\ref{fig:pwave+}.}
\label{fig:pwave-}
\end{figure}

Next, we move on to the scenario of the odd-parity ($p$-wave) spin-septet pair potential for spin-$3/2$ superconductors proposed in Refs.~[\onlinecite{Brydon2016}] and [\onlinecite{Kim2018}]. Hence, we use
\bea
\hat{\Delta}(\boldsymbol{k})&=&\Delta_p\begin{pmatrix}
\frac{3}{4} k_-&\frac{\sqrt{3}}{2} k_z & \frac{\sqrt{3}}{4} k_+&0\\
\frac{\sqrt{3}}{2} k_z& \frac{3}{4} k_+ &0 & -\frac{\sqrt{3}}{4} k_- \\
 \frac{\sqrt{3}}{4}k_+& 0 & -\frac{3}{4}k_-&\frac{\sqrt{3}}{2} k_z\\
 0& -\frac{\sqrt{3}}{4} k_- &\frac{\sqrt{3}}{2} k_z&-\frac{3}{4}k_+
\end{pmatrix},
\label{eq:pwave}
\eea
where $k_{\pm}=k_x\pm ik_y$ and the $p$-wave triplet gap $\Delta_p$ is a real constant. We again calculate the pair amplitude using the anomalous Green's function, Eq.\,\eqref{eq:Green}. We show the real and imaginary parts of the even- and odd-parity pair amplitudes, \ie $\mathcal{\hat{F}}^{Pe}$ and $\mathcal{\hat{F}}^{Po}$, for the $p$-wave pair 
potential in Figs.\,\ref{fig:pwave+} and \ref{fig:pwave-}, respectively. 

From Figs.\,\ref{fig:pwave+}(a-b) we observe that the spin-singlet even-parity pair amplitude is even-$\omega$ 
and has large amplitude for low values of $\omega$, while the spin-triplet even-parity pair amplitudes are zero. 
Four of the spin-quintet pair amplitudes ($q_1$, $q_2$, $q_4$, and $q_5$) are also clearly non-zero, with $q_1= 
q_5$, and all being even-$\omega$ since they are also even-parity, as shown in Figs.~\ref{fig:pwave+}(c-d). 
 Although the $q_3$ pair amplitude is zero in this particular basis, it can be shown to have the same symmetry as 
 the other $q$ pair amplitudes by rotating the basis such that it becomes non-zero. This behavior is the same as 
 what we find for the $s$-wave pair potential in Fig.\ref{fig:swave}.
We also find that four ($s_1$, $s_2$, $s_3$, and $s_5$) out of the seven spin-septet even-parity pair amplitudes are 
finite, but they all have a small magnitude, as shown in Figs.~\ref{fig:pwave+}(e-f). Interestingly, all these spin-septet 
pair amplitudes are odd-$\omega$, as clearly seen in Figs.~\ref{fig:pwave+}(e-f), which are different from all the pair amplitudes discussed so far. 

Moving on to the odd-parity $\mathcal{F}^{Po}$, we find that all the spin-singlet and spin-triplet odd-parity pair 
amplitudes are zero. However, in Figs.\,\ref{fig:pwave-}\,(a-b) we show that the four spin-quintet odd-parity pair 
amplitudes ($q_1$, $q_2$, $q_4$ and $q_5$, with $q_1=q_5$) are finite. They are all odd-$\omega$, unlike the 
spin-quintets even-parity in Fig.\,\ref{fig:pwave+}. These spin-quintet amplitudes are also all comparable in magnitude 
with the heights of the peaks of the pair amplitudes gradually increasing with $\omega$ (including finite outside of 
the plotting window). Here, the $q_3$ pair amplitude is zero, but it can be shown that it follows the same symmetry by rotating the basis. Finally, four of the spin-septet odd-parity pair amplitudes, ($s_1$, $s_2$, $s_3$ and $s_5$) are finite and even-$\omega$, as presented in Figs.~\ref{fig:pwave-}(c-d) and following the pair amplitude symmetry of Eq.~\eqref{eq:pwave}. Other septet amplitudes are zero in our choice of basis, but a rotation of the basis can give rise to non-zero values for the remaining septets ($s_4$, $s_6$ and $s_7$) too, which then all follows the same symmetry classification. Notably, all the non-zero spin-quintet and spin-septet odd-parity pair amplitudes are comparable in magnitude and their symmetries follow the classification shown in Table\,\ref{CPdetails}. 

On the whole, in addition to the pair amplitudes found for the $s$-wave pair potential, we here for $p$-wave 
pair potential find spin-triplet and spin-septet even-parity odd-$\omega$ pair amplitudes as shown in Figs.\,\ref{fig:pwave+} and \ref{fig:pwave-}. These pair amplitudes are in agreement with the classification shown in Table~\ref{CPdetails}, even-parity parts of classes:~1, 5, 6, 8-12, 14 and odd-parity parts of classes:~5, 6, 8-12, 14.

\subsubsection{Chiral even-parity ($d$-wave) pair potential} \label{sec:dwave}
Finally, we consider the chiral even-parity superconducting pair potential proposed in~\cite{Agterberg2017,Boettcher2018},
\begin{align}
\hat{\Delta}(\boldsymbol{k})=\Delta_1\psi_k \eta_s+\Delta
_0(\eta_{xz}+i\eta_{yz}),
\label{eq:dwave}
\end{align}
where
\begin{align}
\eta_{yz} = \frac{1}{\sqrt{3}}(\hat{J}_y \hat{J}_z + \hat{J}_z \hat{J}_y) \,U_T \\
\eta_{xz} = \frac{1}{\sqrt{3}}(\hat{J}_z \hat{J}_x +\hat{J}_x \hat{J}_z) \,U_T
\end{align}
with $\Delta_1$ and $\Delta_0$ being real constants. This pair potential breaks time-reversal symmetry but retain 
inversion symmetry. Here, $\eta_s$ is a spin-singlet state with a form factor $\psi_k$ that is completely isotropic in 
the pairing channel, \ie no mixing between $m=1/2$ and $m=3/2$ channel. The form factor $\psi_k$ breaks time-reversal symmetry but preserve inversion symmetry and is usually taken as $k_z(k_x+ik_y)$. The gap matrix 
$(\eta_{xz}+i \eta_{yz})$ is chiral and transforms under rotation similarly to the spherical harmonics $Y_{2,1}(k)$\,\cite{Agterberg2017}. Instead of point or line nodes as is found generally for a $d$-wave pair potential, this pair 
potential shows an inflated line node, \ie a BFS. In fact, it is the presence of a finite \textcolor{red}{$\Delta_0$} that is responsible for the appearance of the BFS, whereas in absence of $\Delta_0$, there is only a line node in the $k_z=0$ plane and two point nodes on the $k_z$ axis ($k_x=k_y=0$)\,\cite{Agterberg2017}. Equation\,\eqref{eq:dwave} is the most considered pair potential in the literature to explain the appearance of BFS in spin-$3/2$ systems. Thus using this pair potential allows us not just another opportunity to find different pair amplitudes in Table~\ref{CPdetails}, but also to investigate a possible relation between the appearance of BFS and odd-frequency pairing.

\begin{figure}[!thpb]
\centering
\includegraphics[scale=0.695]{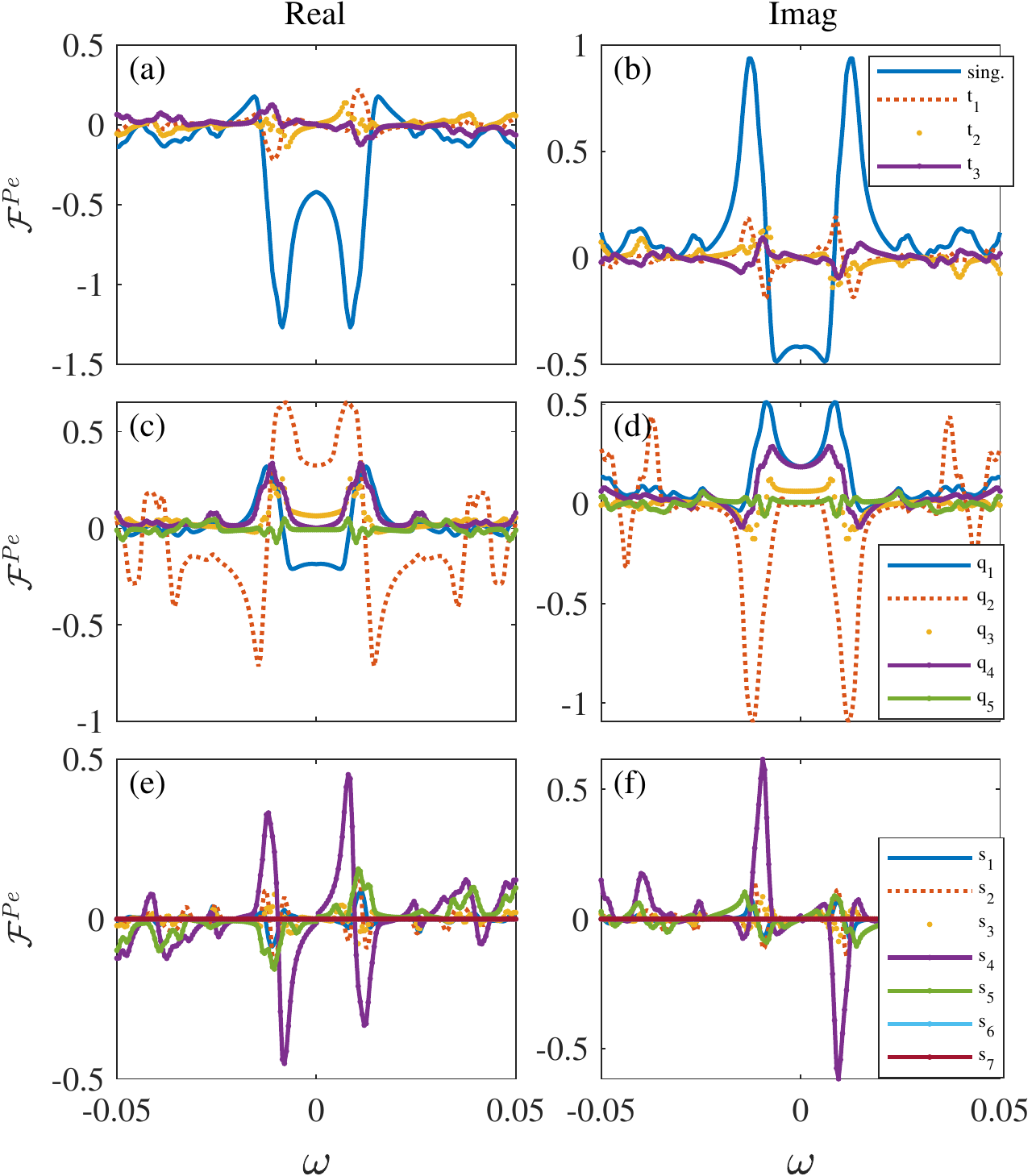}
\caption{Real (a,c,e) and imaginary (b,d,f) part of $\mathcal{F}^{Pe}$ as a function of frequency $\omega$ using 
the pair potential of Eq.\,\eqref{eq:dwave} with $\Delta_0=\Delta_1=0.01$.}
\label{fig:dwave+}
\end{figure}
\begin{figure}[!thpb]
\centering
\includegraphics[scale=0.695]{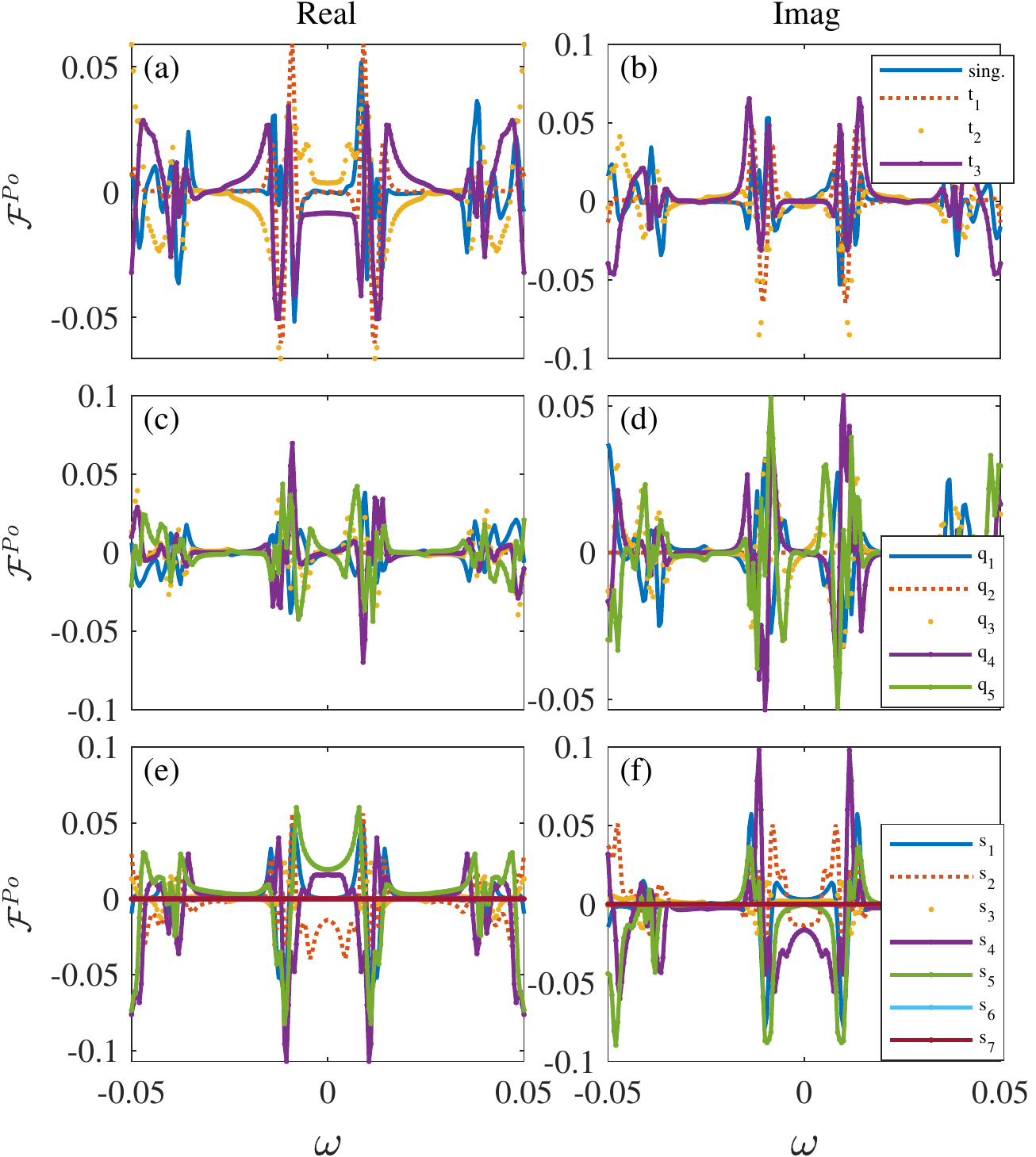}
\caption{Real (a,c,e) and imaginary (b,d,f) part of $\mathcal{F}^{Po}$ as a function of frequency $\omega$. The 
parameter values are same as in Fig.\,\ref{fig:dwave+}. }
\label{fig:dwave-}
\end{figure}
To capture the general behavior of the chiral even-parity pair potential in Eq.~\eqref{eq:dwave}, we set both 
$\Delta_0$ and $\Delta_1$ to non-zero values. We then calculate both the even- and odd-parity spatial parts of $\hat{\mathcal{F}}$, $\mathcal{F}^{Pe}$ and $\mathcal{F}^{Po}$, to find the pair amplitudes and show these in Fig.\,\ref{fig:dwave+} and Fig.\,\ref{fig:dwave-}, respectively. In Figs.\,\ref{fig:dwave+}\,(a-b), we see that the spin-singlet 
even-parity  pair amplitude is finite and even-$\omega$. It is similar to the spin-singlet pair amplitude found for the $s$-wave pair potential in Sec.\,\ref{sec:swave}. However, unlike for $s$-wave pair potential, here all the spin-triplet ($t_1$, $t_2$, and $t_3)$ even-parity pair amplitudes are also finite. Interestingly, these spin-triplet pair amplitudes are odd-$\omega$, following the classification of Table\,\ref{CPdetails}. Moreover, Figs.\,\ref{fig:dwave+}\,(c-d) confirm the even-parity even-$\omega$ behaviors of all the spin-quintet pair amplitudes, in agreement with Table\,\ref{CPdetails}. Some 
of the spin-quintet pair amplitudes are already found for $s$-wave pair potentials. We further find that the spin-septet 
even-parity odd-$\omega$ pair amplitudes are finite and even similar in magnitude to that of the spin-singlet and spin-triplet even-$\omega$ pair amplitudes (except $s_6$ and $s_7$), despite their odd-$\omega$ nature. In particular, 
we here note that the odd-$\omega$ pair amplitudes for the $p$-wave pair potential in Fig.\,\ref{fig:pwave+} are two 
orders of magnitude smaller than what we find here.

We next plot the non-zero odd-parity $\mathcal{F}$ in Fig.~\ref{fig:dwave-}. From both the real and imaginary parts 
of Figs.\,\ref{fig:dwave-}\,(a-b), we see that the spin-singlet pair amplitude is now odd-$\omega$ and all the three 
spin-triplet pair amplitudes, which are even-$\omega$, are finite. This is essentially different from what we find for 
$s$- and $p$-wave pair potentials. Moreover, the spin-quintet odd-parity pair amplitudes are odd-$\omega$, while 
the spin-septet odd-parity pair amplitudes are even-$\omega$. For this chiral pair potential, all the pair amplitudes 
are comparable in magnitude and mostly higher amplitudes are found for the low values of $\omega$. Thus, all the 
pair symmetries revealed for the chiral pair potential in Eq.\,\eqref{eq:dwave} are consistent with Table\,\ref{CPdetails}, 
and particularly, we find finite components in almost all classes 1-14 of Table\,\ref{CPdetails}, with only occasional exceptions where some individual $m$ component is zero. 

Having extracted general results for the $d$-wave chiral pair potential in Eq.~\eqref{eq:dwave}, we note that when $\Delta_0=0$, the even-parity spin-triplet and spin-septet pair amplitudes disappear, as well as the odd-parity spin-singlet and spin-quintet pair amplitudes. As a direct consequence, the $\Delta_0$ pairing term is responsible for generating all the odd-$\omega$ pair amplitudes. As previously established, a finite $\Delta_0$ is also responsible for the appearance of the BFS for this pair potential\,\cite{Agterberg2017}. This establishes numerically a direct relationship between odd-$\omega$ pair amplitudes and BFS for the chiral even-parity pair potential in Eq.~\eqref{eq:dwave}. 
Moreover, in the normal-state Hamiltonian, Eq.~\eqref{ham_N}, the spin-orbit inversion symmetry breaking term $\delta$ 
is the only odd-parity term. Based on symmetry arguments and also confirmed by our numerics, there are no odd-parity pair amplitudes in the absence of this $\delta$ term whenever the superconducting pair potential is even-parity, such as 
for the chiral $d$-wave or the $s$-wave potentials discussed here.

On the whole by studying three different and realistic pair potentials, we show that all different types of the pair symmetries: (1) odd-angular momentum even-parity even-$\omega$, (2) odd-angular momentum odd-parity odd-$\omega$, (3) even-angular momentum even-parity odd-$\omega$, and (4) even-angular momentum odd-parity 
even-$\omega$ symmetry as classified in Table\,\ref{CPdetails}, appear for a generic Hamiltonian describing 
half-Heusler materials. While the classification in Table\,\ref{CPdetails} itself is universal, independent of the pair 
potentials and the other parameter values, our three examples show that all these four classes can easily appear 
in half-Heusler materials. This symmetry classification for spin-$3/2$ superconductors, and particularly the 
identification of odd-$\omega$ pair amplitudes, is one of the main results of this work. In fact, through our examples 
we find all the possible odd-$\omega$ pair amplitudes for spin-$3/2$ superconductors: (i) spin-singlet odd-parity, 
(ii) spin-triplet even-parity, (iii) spin-quintet odd-parity, and (iv) spin-septet even-parity pair amplitudes. Among them, 
we find that the spin-septet even-parity odd-$\omega$ $s_4$ pair amplitude for chiral $d$-wave pair potential is 
largest in magnitude. All the other spin-septet even-parity odd-$\omega$ pair amplitudes and spin-quintet odd-parity 
odd-$\omega$ pair amplitudes found for the $d$-wave pair potential are also large and comparable in magnitude to 
all even-$\omega$ pair amplitude in the same system. In addition to that, also the  spin-quintet odd-parity 
odd-$\omega$ pair amplitudes found for the $p$-wave pair potential are relatively large, only dominated by the 
spin-singlet even-parity even-$\omega$ pair amplitude.

\section{General Analytical Expression for odd-frequency pair amplitude} \label{Anal}
Having numerically found odd-$\omega$ pair amplitudes with all possible spin structures for spin-$3/2$ systems, 
we next derive a general analytical expression for the odd-$\omega$ pair amplitude in these systems. For that, we proceed by rewriting Eq.\,\eqref{eq:Green} using Eq.\,\eqref{eq:bdgH} as 
\begin{align}
 \begin{pmatrix}
{\cal \hat{G}}&{\cal \hat F}\\
\hat {\bar{\mathcal{F}}} &\hat {\bar{\mathcal{G}}}
\end{pmatrix}
= \begin{pmatrix}
i \omega  \hat{\mathcal{I}}  - \hat{H}_0(\boldsymbol{k})& - \hat{\Delta}(\boldsymbol{k}) \\
-\hat{\Delta}(\boldsymbol{k}) & i \omega \hat{\mathcal{I}} + \hat{H}^T_0(\boldsymbol{-k})
 \end{pmatrix}^{-1}
\label{eq:Green2}
\end{align}
We can then write the anomalous Greens function as\,\cite{Triola2020}
\begin{equation}
\hat{\mathcal{F}}=-[(i \omega   \hat{\mathcal{I}} +\hat{H}_0(\boldsymbol{k}))\hat{\Delta}(\boldsymbol{k})^{-1} (i \omega -\hat{H}_0(\boldsymbol{k})) -\hat{\Delta}(\boldsymbol{k})]^{-1}.
\end{equation}
The pair amplitude can finally be rewritten as
\bea
\hat{\mathcal{F}}&=&-\left[(\hat{\Delta}(\boldsymbol{k})-\hat{\Delta}(\boldsymbol{k})^{-1}(\omega^2+\hat{H}_0(\boldsymbol{k})^2) \right. \nonumber \\
&&\left. -\hat{\gamma}^{\prime}\hat{H}_0(\boldsymbol{k})) -i\omega \hat{\gamma}^{\prime}\right] \hat{\mathcal{D}}^{-1}
\label{eq:AnoF}
\eea
where $\hat{\mathcal{D}}=[(\hat{\Delta}(\boldsymbol{k})-\hat{\Delta}(\boldsymbol{k})^{-1}(\omega^2+\hat{H}_0(\boldsymbol{k})^2)-\hat{\gamma}^{\prime} \hat{H}_0(\boldsymbol{k}))^2+\omega^2 \hat{\gamma}^{\prime\,2}]$ and $\hat{\gamma}^{\prime}=\left[\hat{H}_0(\boldsymbol{k}),\hat{\Delta}(\boldsymbol{k})\right]_-$, with the subscript `$-$' denotes the commutation relation. The denominator $\hat{\mathcal{D}}$, being independent of odd powers of $\omega$, 
is an even function of $\omega$. Thus the odd-$\omega$ pair amplitude can be easily identified as
\bea
\mathcal{F}^{T o}=i\omega \hat{\gamma}^{\prime} \hat{\mathcal{D}}^{-1}.
\label{eq:oddF}
\eea
We note that this general expression for the odd-$\omega$ pair amplitude is valid for any system where the pair 
potential $\hat{\Delta}(\boldsymbol{k})$ is a matrix and not a scalar quantity. Thus, Eq.\,\eqref{eq:oddF} holds for 
any multiband system, including the spin-$3/2$ systems studied here. We note that our expression is more general 
than that of Ref.\,[\onlinecite{Triola2020}], where a particular form of multiband Hamiltonian was considered. The 
odd-$\omega$ pair amplitude includes $\hat{\gamma}^{\prime}$, which is the commutator of the normal-state 
Hamiltonian and the pair potential. A similar commutation relation has also been used earlier in the context of multiorbital superconductivity when discussing superconducting fitness~\cite{Ramires2016,Ramires2018}, and also applied for spin-$3/2$ systems~\cite{Kim2021}.To summarize, it is the incompatibility of the diagonal and off-diagonal elements of Eq.\,\eqref{eq:bdgH} that generates odd-frequency pair amplitudes. This incompatibility further indicates that it is necessarily interorbital terms that are responsible for finite odd-$\omega$ pair amplitudes.
This can also be readily confirmed in the results for the $s$-wave pair potential in Section~\ref{sec:swave}. Here, odd-$\omega$ pair amplitudes are completely absent, despite the plethora of different spin channels, as the $s$-wave pair potential commutes with the normal part of the Hamiltonian.

\section{Connection with Bogoliubov Fermi Surface} \label{BFS}
With the above numerical results and general analytical expression for the odd-$\omega$ pair amplitude, we next 
look for whether there exists any general connection between the odd-$\omega$ pair amplitude and the BFS present in spin-$3/2$ systems. In particular, the existence of a BFS leads to a finite DOS around zero energy, thus offering an intriguing connection to odd-$\omega$ pair amplitudes, since a finite zero-energy DOS has previously been used as a characteristic feature of the odd-$\omega$ pairing\,\cite{Yokoyama2007,Linder2010,DiBernardo2015}. 
Moreover, for the chiral $d$-wave pair potential in Eq.~\eqref{eq:dwave} we have already numerically established a direct relationship between the existence of a BFS and finite odd-$\omega$ pair amplitudes. Taken together, this motivates us to look for analytical connections between BFS and odd-$\omega$ pair amplitudes in spin-$3/2$ superconductors.

It has recently been shown that the necessary and sufficient condition for the appearance of BFS in spin-3/2 superconductors is that the time-reversal gap product has to be non-zero\,\cite{Brydon2018}, \ie 
\beq
\hat{\Delta}(\boldsymbol{k}) \hat{\Delta}(\boldsymbol{k})^{\dagger}-\hat{\Delta}_T(\boldsymbol{k}) \hat{\Delta}_T(\boldsymbol{k})^{\dagger}\ne 0. \eeq
Moreover, as discussed in Sec.\,\ref{Anal}, for any spin-$3/2$ superconductor the existence of a finite odd-$\omega$ 
pair amplitude depends on the non-commuting property of the pair potential $\hat{\Delta}(\boldsymbol{k})$ with the normal-state Hamiltonian $\hat{H}_0$ of the superconductor, as expressed in by $\hat{\gamma}'$ in Eq.\,\eqref{eq:oddF}. Thus we can very generally conclude that odd-frequency pairing and BFS can only both be present for a time-reversal symmetry breaking superconductor pair potential that is incompatible with the normal-state Hamiltonian.

Finally, we perform some further analysis to establish a more explicit and analytical connection between odd-$\omega$ pair amplitude and BFS for spin-$3/2$ systems, such as the half-Heusler compounds. 
For this, we only consider the chiral even-parity pair potential of Eq.\,\eqref{eq:dwave} as this particular choice of  pair potential results in a BFS, which is not possible to find using the $s$-wave and $p$-wave pair potentials, and develop an effective low-energy model. To be able to proceed analytically, we first assume that the asymmetric spin-orbit interactions $\gamma,\delta$ in Eq.\,\eqref{ham_N} are zero as they do not play any role in the formation of the BFS. In the absence of these parameters, the energy eigenvalues of the normal-state Hamiltonian of Eq.\,\eqref{ham_N} is doubly degenerate and can be written in the form\,\cite{Brydon2016}
\begin{align}
\varepsilon_\pm({\boldsymbol k})=\varepsilon_0(\boldsymbol{k}) \pm|\boldsymbol{\varepsilon^{\prime}({\boldsymbol{k}}})|
\end{align}
where $\varepsilon_{0}(\boldsymbol{k})=(\alpha+\frac{5\beta}{4})\boldsymbol{k}^2$ and $|\boldsymbol{\varepsilon^{\prime}({\boldsymbol{k}}})|$ is a compact form of writing the magnitude of a five-dimensional vector, $|\boldsymbol{\varepsilon^{\prime}({\boldsymbol{k}}})|=\sqrt{\sum\limits_{i=1}^5\varepsilon_i^2}$ where $\{\varepsilon_1,\varepsilon_2,\varepsilon_3,\varepsilon_4,\varepsilon_5\}=\{\sqrt{3}\beta k_x k_y,\sqrt{3}\beta k_y k_z,\sqrt{3}\beta k_x k_z,\frac{\sqrt{3}\beta}{2}(k_x^2-k_y^2),\beta(k_z^2-\frac{k_x^2+k_y^2}{2})\}$.  
As the chemical potential is usually found in one of the normal-state bands and assuming that the important physical properties of the system are then also coming from that band, we can extract an effective Hamiltonian for this band (here the `$\varepsilon_-$' band). The whole BdG Hamiltonian then simplifies into the form \cite{Agterberg2017, Brydon2018} 
\begin{align}\label{eq:H32min}
\mathcal{\hat{H}}_{b-}(\boldsymbol{k})=\begin{pmatrix}
\tilde\varepsilon({\boldsymbol k})\sigma_0+h_z({\boldsymbol k}) \sigma_z&\psi_-(\boldsymbol{k})i\sigma_y\\
-\psi_-^*(\boldsymbol{k}) i\sigma_y&-\tilde\varepsilon({\boldsymbol k}) \sigma_0-h_z({\boldsymbol k}) \sigma_z,
\end{pmatrix}
\end{align}
where
\begin{align}
\tilde\varepsilon({\boldsymbol k})= \varepsilon_-({\boldsymbol k}) +\zeta({\boldsymbol k}),\\
\psi_-(\boldsymbol{k})=\Delta_1\psi_{\boldsymbol k} - \Delta_0 \frac{\varepsilon_3+i\varepsilon_2}{|\boldsymbol{\varepsilon_{\boldsymbol k}}|}
\end{align}
with
\begin{align}
\zeta ({\boldsymbol k}) &=\frac{|\Delta_0|^2}{4|\boldsymbol{\varepsilon^{\prime}({\boldsymbol k})}|^{3}}(4|\boldsymbol{\varepsilon^{\prime}({\boldsymbol k})}|^2-2\varepsilon_2^2-2\varepsilon_{3}^2),
\nonumber\\
h_z ({\boldsymbol k})&=\frac{|\Delta_0|^2}{|\boldsymbol{\varepsilon^{\prime}({ \boldsymbol k})}|^2}\sqrt{\varepsilon_1^2+\varepsilon_4^2+\varepsilon_5^2}
\label{eq:h}
\end{align}
and energy dispersion given by
\begin{align}
E=\pm h_z({\boldsymbol k}) \pm\sqrt{\tilde\varepsilon({\boldsymbol k})^2+|\psi_-(\boldsymbol k)|^2}.
\end{align}
This is thus a psuedo-spin description of the low-energy physics, which is equivalent to the low-energy basis of the 
original spin-$3/2$ system, where we modified the lower energy band by the effect of the higher energy band. In this 
effective low-energy model, the normal band $\varepsilon_-({\boldsymbol k})$ is modified by the 
$\zeta({{\boldsymbol k}})$ term and a momentum dependent pseudo-magnetic field $h_z({\boldsymbol k})$. Moreover, $\psi_-({\boldsymbol k})$ is the effective nodal superconducting gap potential with a chiral $d$-wave form. It produces a nodal ring on the $k_z=0$-plane, while a finite pseudo-magnetic field $h_z(\boldsymbol{k})$ inflates it into a BFS~\cite{Brydon2018}. 

As the pseudo-magnetic field is in the $z$-direction, Eq.\,\eqref{eq:H32min} can be further simplified by rearranging 
the basis to $(\tilde{c}^\dagger_{\uparrow \boldsymbol{k}}\, \tilde{c}_{\downarrow -\boldsymbol{k}}\,\tilde{c}^\dagger_{\downarrow \boldsymbol{k}}\, \tilde{c}_{\uparrow -\boldsymbol{k}})$, resulting in the BdG Hamiltonian
\begin{align}
{\scriptsize
\mathcal{\hat{H}}_{b-}^{\prime}({\boldsymbol k})=\begin{pmatrix}
\tilde{\varepsilon}_{\boldsymbol k}+h_z({\boldsymbol k}) &\psi_-({\boldsymbol k})&0&0\\
\psi_-({\boldsymbol k}) &-\tilde{\varepsilon}_{\boldsymbol k}+h_z({\boldsymbol k})& 0 & 0\\
0&0 &\tilde{\varepsilon}_{\boldsymbol k}-h_z({\boldsymbol k})&-\psi_-({\boldsymbol k})\\
0 &0&-\psi_-({\boldsymbol k})&-\tilde{\varepsilon}_{\boldsymbol k}-h_z ({\boldsymbol k})
\end{pmatrix}},
\end{align}
where $\tilde{c}^\dagger_{\sigma {\boldsymbol k}}$ is now the creation operator for the pseudo-spin $\sigma$, 
obtained by rotating the original full spin-$3/2$ basis of Eq.\,\eqref{eq:bdgH} into the band basis and then 
projecting the effect of the higher energy bands on the low-energy bands. In this new basis the Green's function 
can be decomposed as 
\beq
{\cal  G}=\begin{pmatrix}
{\cal G}_1&0\\0&{\cal G}_2
\end{pmatrix},
\eeq
where
\bea
{\cal G}_1&=&\frac{1}{(\omega-h_z({\boldsymbol k}))^2-\tilde{\varepsilon}_{\boldsymbol k}^2-\psi_-^2({\boldsymbol k})} \nonumber\\
&&\times \begin{pmatrix}
\omega+\tilde{\varepsilon}_{\boldsymbol k}-h_z({\boldsymbol k})&\psi_-({\boldsymbol k})\\ \psi_-({\boldsymbol k})&\omega-\tilde{\varepsilon}_{\boldsymbol k}-h_z({\boldsymbol k})
\end{pmatrix}
\eea
 with the Green's function of the other block, ${\cal G}_2$, obtained simply by changing $(h_z({\boldsymbol k}),\psi_-({\boldsymbol k})) \rightarrow -(h_z({\boldsymbol k}),\psi_-({\boldsymbol k}))$.
Finally, we can extract the anomalous Green's function as the first block $\mathcal{G}_1$ giving
\begin{align}
{\cal F}_{1}=\psi_-({\boldsymbol k})\frac{\omega^2+h_z^2({\boldsymbol k})-\tilde{\varepsilon}^2_{\boldsymbol k}-\psi_-^2({\boldsymbol k})+ 2\omega h_z({\boldsymbol k})}{(\omega^2+h_z^2({\boldsymbol k})-\tilde{\varepsilon}^2_{\boldsymbol k}-\psi_-^2({\boldsymbol k}))^2- 4\omega^2 h_z^2({\boldsymbol k})}, 
\end{align}
where we now easily find the odd-$\omega$ component as
\bea
\mathcal{F}_{1}^{T o}=\psi_-({\boldsymbol k})\frac{ 2\omega h_z({\boldsymbol k})}{(\omega^2+h_z^2({\boldsymbol k})-\tilde{\varepsilon}^2_{\boldsymbol k}-\psi^2_-({\boldsymbol k}))^2- 4\omega^2 h_z^2({\boldsymbol k})}. \nonumber \\
\eea
This result clearly illustrates how the odd-$\omega$ pair amplitude is directly proportional to the pseudo-magnetic 
field $h_z({\boldsymbol k})$. At the same time, $h_z({\boldsymbol k})$ is also the term explicitly responsible for the appearance of BFS. Thus, for this low-energy effective band model, we find that odd-$\omega$ pairing and BFS always exists together. This also confirms analytically our numerical results in Section \ref{sec:dwave}.

\section{Summary and Conclusions}
\label{conclu}
To summarize, we have introduced the $\mathcal{JPT}=-1$ symmetry classification for superconductivity in 
spin-$3/2$ fermion systems, where the Cooper pairs are composed of two electrons with finite angular momenta 
$j_i = 3/2$. Here, $\mathcal{J}$, $\mathcal{P}$ and $\mathcal{T}$ are the symmetry operators for the 
$z$-component of the total angular momentum, spatial parity, and time (or equivalently, frequency), respectively. 
In addition to spin-singlet and spin-triplet pairing found in conventional spin-$1/2$ systems, there are here two 
additional spin structures namely, spin-quintet and spin-septet Cooper pairs or equivalently pair amplitudes. All 
of these Cooper pairs can further be classified into even- and odd-spatial parity as well as even- and odd-frequency 
pairing. Following the antisymmetry condition $\mathcal{JPT}=-1$, this generates a total of 32 different classes of superconducting pair symmetries. 

To illustrate the existence of the different pair symmetry classes, we have also studied a model suitable to describe 
the superconducting half-Heusler compounds hosting low-energy spin-$3/2$ fermions, using several different superconducting pair potentials suggested in the current literature; even-parity $s$-wave, odd-parity $p$-wave, 
and chiral even-parity $d$-wave pair potentials. By calculating the anomalous Green's function we have numerically accessed the superconducting pair amplitudes for all cases and showed how the different classes can be present, including large odd-$\omega$ pair amplitudes. In particular, we have found that the chiral $d$-wave pair potential generates large spin-septet even-parity odd-$\omega$ pair amplitudes.

We have also derived a general analytical expression for the odd-$\omega$ pair amplitude, applicable to any 
spin-$3/2$ superconductor as well as other multiband systems. Using this expression we have investigated the 
relation between the odd-$\omega$ pair amplitude and the appearance of a Bogoliubov Fermi surface (BFS). 
We have found that both BFS and odd-$\omega$ pairing are always present textcolor{red}{together}for superconducting pair potentials with an odd-gap time-reversal product and non-commuting with the normal-state Hamiltonian. Using a minimal low-energy model of a spin-$3/2$ superconductor \text{red}{with a possible BFS}, we have further been able to sharpen this criteria of co-existence of BFS and odd-$\omega$ pairing and shown that a BFS and a finite odd-$\omega$ pair amplitude are always appearing simultaneously \text{red}{in this effective model}. Based on these results, we speculate that any superconductor with a BFS hosts finite odd-$\omega$ pair amplitudes, albeit the reverse is already known to not be true.

{\it Note added:} During the final stages of preparing this manuscript two other works appeared discussing various
aspects of odd-$\omega$ pairing in spin-$3/2$ superconductors. In Ref.\,[\onlinecite{Kim2021}] a finite odd-$\omega$ 
pair amplitude  was found to favor a $\pi$-state in a Josephson junction consisting of two spin-$3/2$ superconductors. However, this work neither classified the different superconducting pair symmetries, including the different possibilities 
of odd-$\omega$ pair amplitudes, nor considered the relation between odd-$\omega$ pairs and BFSs. In Ref.\,\cite{Miki2021} pairing of the bogolons living on the BFS was shown to be odd-$\omega$ in nature. This is however 
very different from our work as it considers pairing within the BFS, while we consider pairing of the original spin-$3/2$ fermions (which then can generate the BFS). Thus both of these two works are very different from the our results.

\acknowledgments{We thank D.~Chakraborty for technical help and acknowledge financial support from the European Research Council (ERC) under the European Unions Horizon 2020 research and innovation programme (ERC-2017-StG-757553), the Knut and Alice Wallenberg Foundation through the Wallenberg Academy Fellows program, and the Swedish Research Council (Vetenskapsr\aa det Grant No. 2018-03488).}

\begin{appendix}
\section{Matrix operators}\label{j_operators}
The angular momentum operators for spin-$j=3/2$ fermions are expressed in matrix forms as
\begin{align}
\hat{J}_x=\frac{1}{2}\begin{pmatrix}
0& \sqrt{3}&0&0\\
\sqrt{3}&0&2&0\\
0&2&0&\sqrt{3}\\
0&0& \sqrt{3}&0
\end{pmatrix}~~~
\end{align}
\begin{align}
\hat{J}_y=\frac{i}{2}\begin{pmatrix}
0& -\sqrt{3}&0&0\\
\sqrt{3}&0&-2&0\\
0&2&0&-\sqrt{3}\\
0&0& \sqrt{3}&0
\end{pmatrix}\text{, and}
\end{align}
\begin{align}
\hat{J}_z=\frac{1}{2}\begin{pmatrix}
3&0&0&0\\
0&1&0&0\\
0&0&-1&0\\
0&0&0&-3
\end{pmatrix}.
\end{align}
The unitary operator is defined as
 \begin{align}
U_T=\begin{pmatrix}
0&0&0&1\\0&0&-1&0\\0&1&0&0\\-1&0&0&0
\end{pmatrix}.
\end{align}

\end{appendix}

\bibliography{bibfile}

\end{document}